\documentclass{elsarticle}
\usepackage{epsfig}
\usepackage{amssymb}

\def\eq#1{(\ref{#1})} 

\begin{document}

\title{The black-hole area spectrum in the tunneling formalism with GUP}

\author{C.A.S.Silva$^{a,b}$} 
\ead[1]{calex@fisica.ufc.br}
\author{R.R. Landim$^{a}$}
\ead[2]{renan@fisica.ufc.br}

\address{$^{a}$Departamento de F\'{\i}sica, Universidade Federal do
Cear\'a \\ Caixa Postal 6030, CEP 60455-760, Fortaleza, Cear\'a, Brazil} 

\address{$^{b}$Instituto Federal de Educa\c{c}\~{a}o, Ci\^{e}ncia e Tecnologia da Para\'{i}ba (IFPB),\\ Campus Campina Grande.  
\\Rua Tranquilino Coelho Lemos, 671, Jardim Dinam\'{e}rica
I.}

\begin{abstract}
\noindent The black hole area spectrum has been studied in the framework of tunneling mechanism, where a Generalized Uncertainty Principle (GUP) has been 
considered. The results implies in a non evenly spaced spectrum for the black hole area which becomes increasingly spaced as the black hole evaporates.
\end{abstract}

\maketitle

\section{Introduction}

\indent The search for a common description of particle physics and gravity and for a quantum theory
of the gravitational sector is certainly one of the most outstanding and longstanding problems in
physics. Despite important discoveries, at present a reliable theoretical framework is lacking
and even the very meaning of quantum spacetime is not clear.

Hawking discovery that black holes have thermodynamics
properties like entropy and temperature \cite{sw.hawking-cmp43,sw.hawking-prd14,jd.bekenstein-lnc4} has established profound links between black hole physics
and such seemingly very distant fields as thermodynamics, information theory, and quantum theory, and have opened the doors for discussions that have 
improved our understanding about possible the features of a further theory of quantum gravity. In this way, the study of black holes has a significance going far beyond astrophysics.
Actually, black holes may play a major
role in our attempts to shed some light on the nature of a quantum
theory of gravity such as the role played by atoms in the early development
of quantum mechanics.

A consequence of Hawking calculations is the discovery that, due to the instability of the vaccum in the strong gravitational field of a black hole, 
these objects are sources of quantum radiation. In the Hawking's original semiclassical approach, the black hole radiation is continuous. However, several authors
have raised the possibility
that Hawking radiation might in fact have a discrete
spectrum. In its earliest form, this argument traces to Bekenstein's proposal that the
eigenvalues
of the black hole event horizon area are of the form \cite{j.bekenstein-lnc11}

\begin{equation}
A_{n} = \gamma n l_{P}^{2} \label{bek-spectrum}\; ,
\end{equation}

\noindent where $n$ ranges over
positive integers, and $l_{P} = (G\hslash/c^{3})^{1/2}$ is the Planck length.

Diverse computations have been employed in order to find out the correct form of the black hole area spectrum, including some efforts which have been done 
in order to confirm the equally spaced spectrum suggested by Bekenstein \cite{s.hod-prl81, g.kunstatter-prl90, Berti:2004um, Keshet:2007nv, Keshet:2007be, Banerjee:2010be}. 
However, the correct form of the black hole area spectrum remains an open issue.

In this paper we study the black hole area spectrum in the framework of tunneling mechanism, where a Generalized Uncertainty Principle (GUP) has been 
considered. Among
other things, it is a common belief that the Heisenberg uncertainty principle has to be replaced by the so-called generalized uncertainty principle (GUP)
when gravitational interactions are taken
into account. 
Banerjee at al \cite{Banerjee:2010be}, by working with tunneling
formalism, have employed a calculation which have confirmed the equally spaced spectrum proposed by Bekenstein. However this result depends on the use of 
the ordinary Heisenberg Uncertainty Principle, in a way that quantum gravity effects have not been appropriately 
included.

This paper is organized as follows: in section \eq{t-process}, we address the tunneling treatment to black hole evaporation. In section \eq{g-unc-p}, we discuss the
Generalized Uncertainty Principle(GUP) from a micro-black hole gedanken experiment. In section \eq{result-1}, we employ the calculation for the black hole area
spectrum. The section \eq{conclusions}  is devoted to conclusions. In this work we shall consider all fundamental
constants ($c, \hslash, k_{B}, G$) equal to one.

\section{The tunneling process \label{q-tunn}} \label{t-process}

In the region near the horizon, the effective
potential vanishes and there are no grey-body factors. However, the self-consistency of the approach can be seen by recalling that the emission spectrum obtained from
these modes
is purely thermal. This justifies ignoring the grey-body factors. Moreover,
in this region, the theory is dimensionally reduced to a $2$-dimensional theory \cite{Iso:2006ut,
Umetsu:2009ra} whose metric is just the $(t - r)$
sector of the original
metric while the angular part is red-shifted away. Consequently the near-horizon metric has the form

\begin{equation}
ds^{2} = −F(r)dt^{2} + dr^{2}F(r) \label{r-metric} \; .
\end{equation}

\noindent The horizon is defined by the relation $F(r = r_{H}) = 0$ and the surface gravity is given by $\kappa = \frac{F'(r_H)}{2}$ . 

The massless Klein-Gordon
equation $g^{\mu\nu}\nabla_{\mu}\nabla_{\nu}\phi = 0$ under the metric given in \eq{r-metric}, is

\begin{equation}
-\frac{1}{F(r)}\partial_{t}^{2}\phi + F'(r)\partial_{r}\phi + F(r)\partial_{r}^{2}\phi	= 0 \label{k-gordon-eq} \; .
\end{equation}

Taking the standard WKB ansatz $\phi(r, t)=e^{-\frac{i}{\hbar}S(r, t)}$ and substituting the expansion for $S(r, t)$

\begin{equation}
S(r,t) = S_{0}(r,t) + \sum_{i=1}^{\infty}h^{i}S_{i}(r,t)
\end{equation}

\noindent in \eq{k-gordon-eq}, we obtain the solutions for $\phi$ in the semiclassical limit \cite{Banerjee:2008sn, Banerjee:2009wb}

\begin{eqnarray}
\phi_{in}^{(R)} &=& e^{-i\omega u_{in}} \;\;\; ; \;\;\; \phi_{in}^{(L)} = e^{-i\omega v_{in}} \nonumber \\
\phi_{out}^{(R)} &=& e^{-i\omega u_{out}} \;\;\; ; \;\;\; \phi_{out}^{(L)} = e^{-i\omega v_{out}} \label{f-in-out}
\end{eqnarray}

\noindent where the quantity $\omega$ is the energy of the particle as measured by an asymptotic observer. Here ``$R(L)$" refers to the outgoing (ingoing) mode, while
``in(out)" stands for inside (outside) the event horizon. The null coordinates $(u, v)$ are defined as

\begin{eqnarray}
u &=& t − r^{*} ; \nonumber \\
v &=& t + r^{*} ; \nonumber \\ 
dr^{*} &=& drF(r). \label{uv-eq}
\end{eqnarray}

In the context of the tunneling formalism, a virtual pair of particles is produced in the black hole. One member of this pair can quantum mechanically tunnel through
the horizon. This particle is observed at infinity while the other goes towards the center of the black hole. While crossing the horizon the nature of the coordinates
changes. This can be accounted by working with Kruskal coordinates which are viable in both sectors of the black-hole event horizon. The Kruskal time (T) and space (X)
coordinates inside and outside the horizon are defined as \cite{a.raychaudhuri-grac}

\begin{eqnarray}
T_{in} &=& e^{\kappa r∗_{in}} cosh(\kappa t_{in}) ; \;\;\;\; X_{in} = e^{\kappa r∗_{in}} sinh(\kappa t_{in}) \; ;\nonumber \\
T_{out} &=& e^{\kappa r∗_{out}} sinh(\kappa t_{out}) ; \;\;\; X_{out} = e^{\kappa r∗_{out}} cosh(\kappa t_{out}) \; .
\end{eqnarray}

These two sets of coordinates are connected through the following relations:

\begin{eqnarray}
t_{in} &=& t_{out} - i\frac{\pi}{2\kappa}\; , \nonumber \\
r_{in} &=& r_{out} + i\frac{\pi}{2\kappa} \label{t-r-in} \; .
\end{eqnarray}

In this way, the Kruskal coordinates get identified as 

\begin{eqnarray}
T_{in} &=& T_{out}\; , \nonumber \\
X_{in} &=& X_{out}.
\end{eqnarray}

Employing equations \eq{t-r-in} in equation \eq{uv-eq}, we can obtain the
relations that connect the radial null coordinates defined inside and outside the black-hole event horizon

\begin{eqnarray}
u_{in} &=& t_{in} - r_{in} = u_{out} - i\pi/\kappa \; ,\nonumber \\
v_{in} &=& t_{in} + r_{in} = v_{out} .
\end{eqnarray}

Under these transformations the modes in equations \eq{f-in-out} which are traveling in the ``in" and ``out" sectors of the black-hole horizon are connected through the
expressions

\begin{eqnarray}
\phi_{in}^{(R)} &=& e^{-\frac{\pi\omega}{\kappa}}\phi_{out}^{(R)} \; ,\nonumber \\
\phi_{in}^{(L)} &=& \phi_{out}^{(L)} \label{f-in-out-2}.
\end{eqnarray}

Concentrating on the modes located inside the horizon, the L mode is trapped while the R mode tunnels through the horizon \cite{Banerjee:2008sn, a.raychaudhuri-grac}.
The
probability for the R mode to
travel from the inside to the outside of the black hole, as measured by an external observer, is given by

\begin{equation}
P^{R} = \mid \phi_{in}^{(R)} \mid ^{2} = \mid e^{-\frac{\pi\omega}{\kappa}}\phi_{out}^{(R)} \mid^{2} = e^{\frac{2\pi\omega}{\kappa}}\;\; ,
\end{equation}

\noindent where equation \eq{f-in-out-2} has been used to extract the final expression. Since the measurement is done from the outside, $\phi_{in}(R)$ has to be
expressed in terms of
$\phi_{out}^{(R)}$. Therefore the average value of the energy, measured from outside, is written as

\begin{equation}
\langle \omega \rangle = \frac{\int_{0}^{\infty}d \omega\; \omega P'^{R}}{\int_{0}^{\infty}d \omega\; P'^{R}} = T_{H}.  \label{mv-omega}
\end{equation}

\noindent where $T_{H} = \frac{\kappa}{2\pi}$ is the Hawking temperature. In a similar way, one can compute the average squared energy of the particle, detected
by an asymptotic observer,

\begin{equation}
\langle \omega^{2} \rangle = \frac{\int_{0}^{\infty}d \omega\; \omega^{2} P'^{R}}{\int_{0}^{\infty}d \omega\; P'^{R}} = 2T_{H}^{2}. \label{mv-omega-2}
\end{equation}

\noindent Hence it is straightforward to evaluate the uncertainty in the detected energy $\omega$ by combining equations \eq{mv-omega} and \eq{mv-omega-2},

\begin{equation}
(\Delta \omega ) = \sqrt{\langle \omega^{2} \rangle -\langle \omega \rangle^{2}}  = T_{H}\;\; , \label{mv-delta-omega}
\end{equation}

\noindent which is nothing but the Hawking temperature $T_{H}$.

\section{The generalized uncertainty principle} \label{g-unc-p}

The generalized uncertainty principle arises from the Heisenberg uncertainty
principle when gravity is taken into account.
In this section, we will derive a GUP via a micro black hole
gedanken experiment, following closely the content of
\cite{Scardigli:1999jh}.
When we measure a position with precision of order $\Delta X$,
we expect quantum fluctuations of the metric field around the
measured position with energy amplitude

\begin{equation}
\Delta \omega \sim \frac{1}{2 \Delta X}.
\end{equation}

\vspace{5mm}

\noindent The Schwarzschild radius associated with the energy $\Delta \omega$

\begin{equation}
R_{S} = 2\Delta \omega
\end{equation}

\noindent falls well inside the interval $\Delta x$ for practical cases. However,
if we want to improve the precision indefinitely, the fluctuation
$\Delta \omega$ would grow up and the corresponding $R_{S}$ would
become larger and larger, until it reaches the same size as $\Delta X$.
As it is well known, the critical length is the Planck length, and the associated energy is the Planck energy $\varepsilon_{P}$.

If we tried to further decrease $\Delta X$, we should concentrate in
that region an energy greater than the Planck energy, and this
would enlarge further the Schwarzschild radius $R_{S}$, hiding more and more details of the region beyond the event horizon
of the micro hole. The situation can be summarized by the
inequalities

$$\Delta X = \left\{\begin{array}{rc}
\frac{1}{2 \Delta \omega}&\mbox{for}\quad \Delta \omega \leq \varepsilon_{P}\; , \\ \\
 2 \Delta \omega &\mbox{for} \quad \Delta \omega > \varepsilon_{P} \;,
\end{array}\right. 
$$

\vspace{5mm}

\noindent which, if combined linearly, yield

\begin{equation}
\Delta X \geq \frac{1}{2 \Delta \omega} + 2 \Delta \omega\; . \label{gup}
\end{equation}

\vspace{5mm}
This is a generalization of the uncertainty principle to cases
in which gravity is important, i.e. to energies of the order of
$\varepsilon_{P}$. We note that the minimum value of $\Delta X$ is reached for
$(\Delta \omega)_{min} = \varepsilon_{P}$ and is given by $(\Delta X)_{min} = 2 l_{P}$.

\section{The black hole area spectrum in the tunneling formalism with GUP} \label{result-1}

In this section, we will derive the black hole area spectrum using the generalized uncertainty process in the quantum tunneling formalism addressed in the section
\eq{q-tunn}.

According to Bekenstein \cite{Bekenstein:1974ax}, the change in area of a black hole caused either by an absorption or by an emission of a particle is given by

\begin{equation}
\Delta A \geq 8 \pi \int_{V} x T_{00} dV \;\; , \label{delta-a}
\end{equation}

\noindent where $x$ is the distance of the center of mass of the particle from the horizon and $T_{00}$ represents the energy density corresponding to the particle.
Here
$V$ stands for the volume (a $3$-surface) of the system, i.e. the black hole and the particle, outside the black hole, at a constant time.

Following \cite{Banerjee:2010be}, we can consider, based on dimensional grounds,
the position x of the emitted particle to be of the order of the uncertainty in particle's position, i.e. $(\Delta X)$. Then one can set $x =
\epsilon\Delta X$. Therefore, equation \eq{delta-a} can be written as

\begin{equation}
\Delta A \geq 8 \pi \Delta X \int_{V} T_{00} dV \;\; . \label{delta-a2}
\end{equation}

It is evident that the value of the integration on the right-hand side of equation \eq{delta-a} is exactly the energy of the outgoing particle. Since the integration is
performed at
a constant time over the whole space outside the black hole, it is legitimate to identify the energy of the particle as computed through the integration with the
average energy of the particle given by equation \eq{mv-omega}. 

\begin{equation}
\Delta A \geq 8 \pi \epsilon \Delta X \Delta \omega \;\; ,
\end{equation}

Using the equations \eq{mv-delta-omega} and \eq{gup}, we have that

\begin{equation}
\Delta A \geq 8 \pi \epsilon \Big( \frac{1}{2} + 2 T_{H}^{2} \Big)\; .
\end{equation}

\noindent In a way that

\begin{equation}
\Delta A_{min} = 8 \pi \epsilon \Big( \frac{1}{2} + 2 T_{H}^{2} \Big)\; . \label{area-spac}
\end{equation}

The equation  \eq{area-spac} give us an area spectrum which is not equally spaced, but increases as the black hole evaporates, in a different way from 
the Bekenstein proposal.

\section{Conclusions} \label{conclusions}

The black hole area spectrum has been studied in the framework of tunneling mechanism, and in the presence of a Generalized Uncertainty Principle (GUP). We get the
expression for the change in the black-hole area in this framework. The results implies in a non evenly spaced spectrum for the black hole area in a different way
from the Bekenstein proposal. In this case, the lines of the area spectrum becomes more and more spaced
 as the black hole evaporates.

\expandafter\ifx\csname url\endcsname\relax \global\long\def\url#1{\texttt{#1}}
\fi \expandafter\ifx\csname urlprefix\endcsname\relax\global\long\def\urlprefix{URL }
\fi


\begin{thebibliography}{23}

\bibitem[1]{sw.hawking-cmp43} S.W.~Hawking, Particle Creation by Black
Holes, Commun.Math.Phys. 43 (1975) 199--220.

\bibitem[2]{sw.hawking-prd14} S.W.~Hawking, Breakdown of Predictability
in Gravitational Collapse, Phys.Rev.D 14 (1976) 2460--2473.

\bibitem[3]{jd.bekenstein-lnc4} J.D.~Bekenstein, Black holes and
the second law, Lett. Nuovo Cim. 4 (1972) 737--740. 

%
\bibitem[4]{j.bekenstein-lnc11}
   J.D. Bekenstein, Lett. Nuovo Cimento 11 (1974) 467.




\bibitem[5]{s.hod-prl81}   
  S.~Hod,
     Phys. Rev.  Lett. 81, (1998) 4293.
  

\bibitem[6]{g.kunstatter-prl90}
  G.~Kunstatter,
     Phys. Rev.  Lett. 90, (2003) 161301.
 


\bibitem[7]{Berti:2004um}
  E.~Berti, V.~Cardoso and S.~Yoshida,
  Highly damped quasinormal modes of Kerr black holes: A Complete numerical
  investigation,''
  Phys.\ Rev.\  D {\bf 69}, 124018 (2004).
  



\bibitem[8]{Keshet:2007nv}
  U.~Keshet and S.~Hod,
  Analytic study of rotating black-hole quasinormal modes,
  Phys.\ Rev.\  D {\bf 76}, 061501 (2007)
  [arXiv:0705.1179 [gr-qc]].
  
%
%

\bibitem[9]{Keshet:2007be} 
  U.~Keshet and A.~Neitzke,
  Asymptotic spectroscopy of rotating black holes,
  Phys.\ Rev.\  D {\bf 78}, 044006 (2008).
  
%
%
%
%
%
%
%
%
%
%
%


\bibitem[10]{Banerjee:2010be}
  R.~Banerjee, B.~R.~Majhi and E.~C.~Vagenas,
  A Note on the Lower Bound of Black Hole Area Change in Tunneling
  Formalism,''
  Europhys.\ Lett.\  {\bf 92}, 20001 (2010).
 
 

 

\bibitem[11]{Iso:2006ut}
  S.~Iso, H.~Umetsu and F.~Wilczek,
  Anomalies, Hawking radiations and regularity in rotating black holes,
  Phys.\ Rev.\  D {\bf 74}, 044017 (2006).
 
 


\bibitem[12]{Umetsu:2009ra} 
  K.~Umetsu,
  Hawking Radiation from Kerr-Newman Black Hole and Tunneling Mechanism,
  Int.\ J.\ Mod.\ Phys.\  A {\bf 25}, 4123 (2010).
  
  




\bibitem[13]{Banerjee:2008sn}
  R.~Banerjee and B.~R.~Majhi,
   ``Connecting anomaly and tunneling methods for Hawking effect through
  chirality,''
  Phys.\ Rev.\  D {\bf 79}, 064024 (2009).
 
  


\bibitem[14]{Banerjee:2009wb} 
  R.~Banerjee and B.~R.~Majhi,
  Hawking black body spectrum from tunneling mechanism,
  Phys.\ Lett.\  B {\bf 675}, 243 (2009).
  
  


\bibitem[15]{a.raychaudhuri-grac} 
  A.K. ~Raychaudhuri S.~Banergi and R.~Banerjee, 
General Relativity, Astrophysics, and Cosmology (Springer, New York).


\bibitem[16]{Scardigli:1999jh}
   F.~Scardigli,
   Generalized uncertainty principle in quantum gravity from micro - black
   hole Gedanken experiment,''
   Phys.\ Lett.\  B {\bf 452}, 39 (1999)
   

\bibitem[17]{Bekenstein:1974ax}
  J.~D.~Bekenstein,
  Generalized second law of thermodynamics in black hole physics,
  Phys.\ Rev.\  D {\bf 9}, 3292 (1974).
 






\end{thebibliography}
\end{document}